 \documentclass[11pt,preprint2]{aastex}

 \newcommand{\ee}{{$\delta$ Scuti }}
 \newcommand{\aaa}{\hbox{M$_{\odot}$}}
 
 \def\hr{\hbox{$^{\rm h}$}}                 
 \def\deg{\hbox{$^\circ$}}                  
 \def\fdeg{\hbox{$.\!\!^\circ$}}            
 \def\fsec{\hbox{$.\!\!^{\rm s}$}}          
 \def\arcm{\hbox{$^\prime$}}                
 \def\arcs{\hbox{$^{\prime\prime}$}}        
 \def\farcs{\hbox{$.\!\!^{\prime\prime}$}}  
 \def\mm{\hbox{$^{\rm m}$}}                 
 \def\fmm{\hbox{$.\!\!^{\rm m}$}}           

\shorttitle{Cluster IC 348}
\shortauthors{K{\i}z{\i}lo\u{g}lu et al.}

\begin{document}

 \title{ROTSE observations of the young cluster IC 348 }

 \author{\"U. K{\i}z{\i}lo\u{g}lu, N. K{\i}z{\i}lo\u{g}lu\altaffilmark{1}, A. Baykal }
\affil{ Physics Department, Middle East Technical University ,Ankara 06531,
Turkey}
\altaffiltext{1}{Proofs to:  N. K{\i}z{\i}lo\u{g}lu  Physics Dept.Middle East
Tech. Unv. 06531 Ankara, Turkey. \\ e-mail:nil@astroa.physics.metu.edu.tr}

\begin{abstract}
CCD observations of stars in the young cluster IC 348 were obtained from 
2004 August to 2005 January with a 0.45 m ROTSEIIId robotic reflecting
 telescope at the Turkish National Observatory site, Bak{\i}rl{\i}tepe, Turkey.
The timing analysis of selected stars whose X-Ray counterpart were detected by 
Chandra X-Ray Observatory 
were studied. The time series of stars were searched for rotational 
periodicity by using different period search methods. 35 stars were found 
to be periodic with periods ranging from 0.74 to 32.3 days.
Eighteen of the 35 periodic stars were new detections. Four of the new
detections were CTTSs and the others were WTTSs and G type (or unknown
spectral class) stars. In this study, we confirmed the stability of rotation
periods of TTauri stars. The periods obtained by Cohen et al. and us
were different by 1$\%$. We also confirmed the 3.24 h pulsation period of 
H254 which is a \ee type star as noted by Ripepi et al. but the
other periods detected by them were not found. 
We examined correlation
 between X-ray luminosity and rotational period of our sample of TTSs.
There is a decline in the rotational period with X-ray luminosity for 
late type TTSs.
\end{abstract}



\newcommand{\xr}[1]{\parindent=0.0cm\hangindent=1cm\hangafter=1\indent#1\par}

\section{Introduction}
  
IC 348 is a young (less than 10Myr) and nearby cluster ( distance 316 pcs)
located in the Perseus complex (Lada $\&$ Lada 1995, Trullols $\&$
 Jordi 1997, Herbig 1998, Luhman et al. 1998).
This cluster  has a number of T Tauri stars
(Herbig 1954) which are lower mass (lower than 1.5 \aaa) Pre-Main-Sequence
stars. Deep near infrared imaging survey of IC 348 in the J, H and K bands by
Lada $\&$ Lada (1995) resulted with 380 NIR sources as probable cluster 
members. Herbig (1998) made a survey for stars having H$_\alpha$ emission
and discovered over 110 emission line stars brighter than R= 19 magnitude. 
He found the proportion of WTTSs (weak line TTSs with H$_\alpha$ 
equivalent width  below 10\AA~  and H$_\alpha$ emission can be assumed
 to be chromospheric origin) to CTTSs (classical TTSs with H$_\alpha$
equivalent width above 10\AA~ and H$_\alpha$ emission is probably dominated by
the accretion of circumstellar material on to the stars)
 as 58:51. CTTSs exhibit infrared excess and 
show a varying photometric light curves irregularly.
 WTTSs show spectroscopic and photometric periodic variability
on time scales of days caused by rotational modulation due to magnetic
 activity. 
Luhman et al. (1998) performed deep infrared and optical spectroscopy of IC 348
 and found that nearly 25$\%$ of stars within the core of IC 348 and younger
 than 3 Myr exhibits signature of disks in the form of strong H$_\alpha$.

Herbst et al. (2000) studied the photometry of 150 stars and discovered 
19 periodic variables with periods ranging from 2.24 to
16.2 days and masses ranging from 0.35 to 1.1 \aaa.
This variability is caused by the rotation of the surface with large cool spots whose pattern is often stable for many rotation periods.
Recently Cohen et al. (2004) presented results based on 5 yr of monitoring this cluster and found that these periodic stars show modulations of
 their amplitude, mean brightness and light curve shape on time scales of less than one year.

X-ray observations of IC 348 with ROSAT by Preibisch et al. (1996) resulted with detection of 116 X-ray sources. They found probable new cluster members.
They suggested that these were presumably weak line T Tauri stars
because of their X-ray properties. WTTSs seem to be stronger X-ray emitters
than the CTTSs. They could not find any significant correlation
between the H$_\alpha$ luminosity and X-ray luminosity indicating that 
H$_\alpha$ emission is not a chromospheric emission for CTTSs.
Preibisch $\&$ Zinnecker (2001) detected 215 X-ray sources with the Advanced
CCD Imaging Spectrometer on board the Chandra X-Ray Observatory.
58 of these sources were identified as new cluster members.
They did not find significant differences between the X-ray properties of WTTSs and CTTSs. About 80$\%$ of cluster members with masses between 0.15 and 2 \aaa
were identified as visible X-ray sources. The observed X-ray emission
 was explained as coronal emission for WTTSs. Chandra X-ray detection
fraction of the IC 348 cluster was high for spectral types between
the late F and M4. In their next study, Preibisch $\&$ Zinnecker (2002)
found a tight correlation between X-ray luminosity and H$_\alpha$
luminosity for the WTTSs. They suggested that the chromosphere was heated by
 X-rays from the overlying corona. The CTTSs did not show such a 
relation since H$_\alpha$ emission comes mainly from accretion  processes.
They pointed out that the use of H$_\alpha$ emission as an indicator
for circumstellar material had some problems. 

The main goal of this study is to find out if there is a variability in
 the light curve of some cluster members which have X-ray counterparts.
We wanted to examine correlation between X-ray
luminosity and rotational period  of stars in this cluster
in order to see whether rotation is an important 
parameter governing the X-ray emission. We chose
 some X-ray emission sources which were detected and located by Chandra X-Ray
Observatory. We investigated the corresponding optical 
 light curves of these sources obtained by robotic ROTSEIIId 
telescope in order to search for variability.
We also wanted to  investigate the observational results for \ee  
star H254 which is a member of this cluster. 
This star was previously detected by Ripepi et al. (2002)
and found as a \ee star.
 On the basis of observations of Luhman et al. (1998) H254 (L= 31.4
 L$_{\odot}$ and T$_{e}$= 7200 K) is 
 inside the theoretical pulsational
instability strip for the PMS stars determined by Marconi and Palla (1998).
Ripepi et al. obtained that H254 pulsates with a pulsation frequency of
7.406 $d^{-1}$. This frequency was confirmed in our observations.
Ripepi et al.  calculated that this star pulsates either in the fundamental
mode or in the first overtone. They gave a mass range of 2.3 and 2.6 \aaa
for this star by computing a sequence of linear non-adiabatic models.
 In the second section the observations and
data reduction were discussed. The results  and discussion 
related with the periodic
 variables in this cluster were given in the 
section 3. We summarized
our results in the last section.

\section{Observations and Data reduction}

The CCD observations of cluster stars were performed during August, 2004 and 
January, 2005 with ROTSEIIId robotic reflecting telescope 
located at the Turkish National Observatory (TUG) site, 
Bak{\i}rl{\i}tepe, Turkey. ROTSEIII telescopes were described
 in detail by Akerlof et al. (2003). They were designed
for fast ($\sim$6 s) responses to Gamma-Ray Burst triggers from
satellites such as Swift.
 It has a 45 cm mirror and operates without filters.
It has equipped with a CCD, 2048$\times$2048 pixel,
 the pixel scale is 3.3 arcsec
per pixel for a total field of view
 1.$^{\circ}$85$\times$1.$^{\circ}$85. A total of about 1800 CCD
 frames were collected during the observations. Due to
 the other scheduled observations and atmospheric conditions
 we obtained 3 - 40 frames at each night
with an exposure time of 5 sec. 
All images were automatically dark- and flat-field corrected
as soon as they were exposed.
For each corrected image aperture photometry by SExtractor package 
(Bertin $\&$ 
Arnouts 1996) were applied using an aperture of 5 pixels in diameter
to obtain the instrumental magnitudes. Then these magnitudes were calibrated by comparing all the field stars against USNO A2.0 R-band catalog with a 
triangle-matching technique.
Barycentric corrections were made to the times of 
each observation by using JPL DE200 ephemerides prior to the timing analysis 
with the period determination methods.   

\section{Results and Discussion}

\subsection{Pulsation period of \ee star H254}

\begin{figure}[h]
 \epsscale{0.2}
 \includegraphics[clip=true,scale=0.3,angle=270]{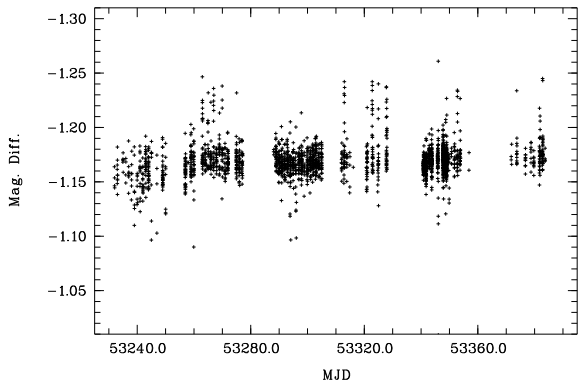}
 \figcaption{RotseIIId light curve of H254. Error bars on data points 
             are not shown for clarity however estimated errors are
             of the order $\sim$0\fmm02}\label{fig1}
 \end{figure}

We first attempted to determine the pulsation period of star H254
(spectral type F0, Harris et al. 1954) 
($\alpha$=03\hr 44\mm 31\fsec2, $\delta$=+32\deg 06\arcm 22\farcs1)
using our nearly 150 days of
 observational data. Ripepi et al. (2002) identified four frequencies for
 this source by using their eleven days  observations.
 One of these frequencies was at 7.406 $d^{-1}$
 which is typical of \ee type pulsators. They
also reported three more frequencies and explained that these were resulted
from the long term behavior associated with a daily variation of H254
 and partially, with the similar variability in their comparison star H20.
We used differential magnitudes which reduces the systematic effects
since we are interested in the time series analysis. As a comparison star
we chose H89  (see section 3.2) 
($\alpha$=03\hr 44\mm 21\fsec0, $\delta$=+32\deg 07\arcm 38\farcs7)
which has a spectral type F8.
Figure 1 shows ROTSEIIId light curve 
 ($\delta$m$_{R}$=m$_{R}$$^{254}$-m$_{R}$$^{89}$) 
obtained between the nights of MJD 53232 and MJD 53382.
 Period of variation in this light curve was determined by using three separate numerical period searching routines. One is Period98 
(by Sperl: available at
www.astro.univie.ac.at/$^{\sim}$dsn/). The other two are the method of 
Scargle (Scargle 1982) and the Clean method (Roberts et al. 1987).
These periodograms are essentially discrete Fourier transform of the input 
time series.
To search any periodicity in the  differential light curves, 
we applied these different period search algorithms mentioned above.
 \placefigure{fig1}
 For analyzing the periodicity in the light curves periodogram provides an
 approximation to the power spectrum. To be sure about the periodicity
 we applied these different period finding methods.

\begin{figure}[h]
 \includegraphics[clip=true,scale=0.30,angle=270]{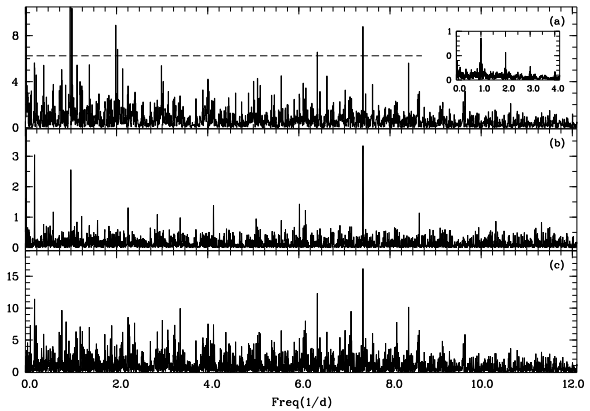}
 \figcaption{Power spectra for H254. Panel (a): Scargle algoritm, 
             (b): Clean algoritm and (c): Period98. Dotted line
             on the upper panel represents $3\sigma$ confidence level. 
             Inset is the spectrum of the window function.}\label{fig2}
 \end{figure}

Figure 2 shows the amplitude and 
power spectra of H254. All of 
them displays the frequency 7.406 $d^{-1}$.
The inset in the first panel shows the window function 
which is used to describe the response of data analysis system to a perfect 
sine wave. The peak at one day (and its harmonics) in the periodograms is 
a signature of nightly windowing of the sampling frequency.
 \placefigure{fig2}

\begin{figure}[h]
 \includegraphics[clip=true,scale=0.3,angle=270]{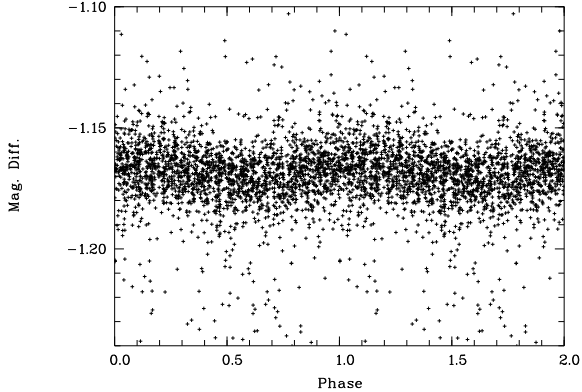}
 \figcaption{Light curve of H254 phased with the frequency
            7.406 d$^{-1}$. }\label{fig3}
 \end{figure}

For the statistics of periodograms we employed the method of Scargle (Scargle
1982) and evaluated the confidence levels of periodicities. 
We estimated the noise level of the periodogram by fitting a constant line.
The probability of a signal above this level has an exponential
probability distribution 
\[
  1- P(Z)=  (1- e^{Z})^{N}  
\]
which is essentially a $\chi^{2}$ distribution for two degrees of freedom.
$Z$ is the power at a given frequency and $N$ is the number of
frequencies sampled. For given parameters the confidence level of the signal 
was found. The confidence level of the signal for the maximum power 
at 7.406 $d^{-1}$ is 
 more than $ 5\sigma $ level signal detection.
 As seen from Figure 2a
all other detected powers are below the $3\sigma $ detection level
which indicate that 0.157, 0.283 and 0.931 $d^{-1}$ frequencies
 detected by Ripepi et al. (2002)
are not present in our light curve. The light curve phased with the frequency
 7.406 $d^{-1}$ is shown in Figure 3. The amplitude of pulsation is 4.1 mmag
which is comparable with V band amplitude (5.4 mmag) given by Ripepi et al.
 \placefigure{fig3}

\begin{figure}[h]
 \includegraphics[clip=true,scale=0.3,angle=270]{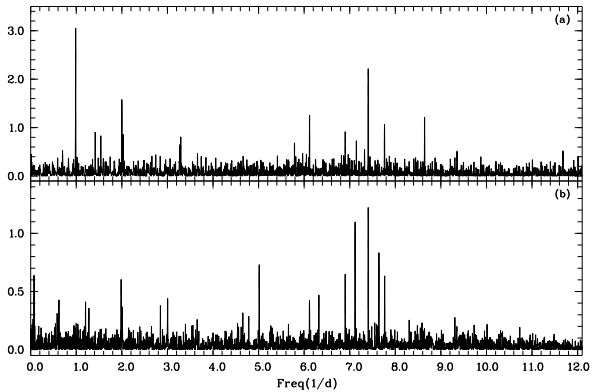}
 \figcaption{Amplitude spectrum of H254 with other set of reference stars 
               (a) H261 (b) H20.}\label{fig4}
 \end{figure}

The same period was found using two more different  comparison stars,
 H261 (spectral type F2, 
 $\alpha$=03\hr 44\mm 24\fsec6, $\delta$=+32\deg 10\arcm 14\farcs4)
 and  H20 (spectral type F8, 
 $\alpha$=03\hr 43\arcm 58\fsec1, $\delta$=+32\deg 09\arcm 47\farcs5).
The amplitude spectra which were obtained
by using Clean method are shown in Figure 4. The 
 peak at the frequency 7.406 $d^{-1}$  corresponding
 to a period of around 3.24 h 
  is seen clearly.
 \placefigure{fig4}

\subsection{Periodic variations in Optical Counterparts of Chandra Sources}

In this part of the study we searched for the timing properties 
of the optical counterparts of selected Chandra X-Ray sources. 
X-ray images of the cluster IC 348 with the Advanced CCD Imaging Spectrometer 
on board the Chandra X-Ray observatory were studied by Preibisch $\&$ Zinneker
(2001). They determined the positions
 and count rates of the 215 individual X-ray
 sources. Identification of the
optical counterparts of the X-ray sources
with masses 0.15 and 2 \aaa were performed for 161 X-ray sources.

\begin{figure}[h]
 \includegraphics[clip=true,scale=0.3,angle=270]{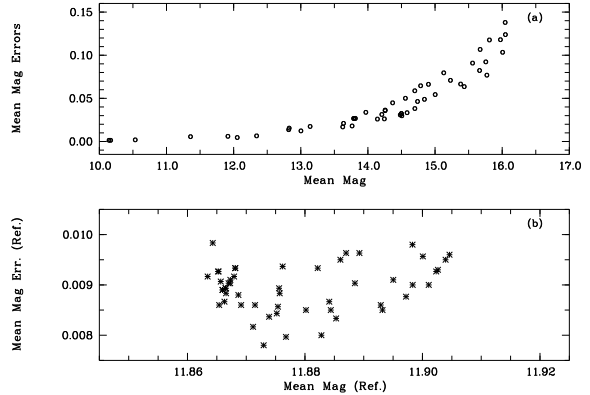}
 \figcaption{Upper panel shows mean magnitude errors
     in calculating the mean magnitudes  for each star in our sample. 
     The mean magnitude of reference stars and their mean errors 
     for each star under consideration are shown in the lower panel.
     }\label{fig5}
 \end{figure}

The positions of Chandra sources whose optical counterparts were
identified by Preibisch $\&$ Zinneker (2001) 
were cross correlated with the positions of
ROTSE objects.
The main criteria for the selection is 3\farcs3/pixel resolution of the ROTSE
CCD frames. Hence, to match a known coordinate 3 pixel (10\arcs) diameter
aperture is used. Secondly, if there is an object closer than 4 pixels 
it is rejected. 
The exposure time is 5 seconds for each frame. This allows us to
observe most of the bright stars of IC 348 without overexposing the frames.
With this exposure time stars with  magnitudes between 10 and 14 are 
well detected and it is also possible to detect stars upto 17th magnitude
depending on atmospheric conditions.  
For each frame mean FWHM of the point spread function (PSF) is calculated 
for stars with 
 10 $>$ m$_{R}$ $>$ 14
 and if the mean FWHM $>$ 2 pixel (6\farcs6) that frame 
is also rejected. 
Figure 5a shows the mean of the magnitude measurement  errors 
assigned to each star in finding their mean magnitudes 
during the whole observation period.
For fainter stars the magnitude determining accuracy decreases.
The magnitude errors should be excluded from the measured variations
in order to obtain the correct intrinsic variabilities. The lower limit for
the systematic measurement errors is 0.002 for the brightest star
in our figure. For magnitudes of stars $\sim$16 mag, this error
is about 0.15 mag. As this error increases with increasing magnitude
the measurement of intrinsic variability becomes difficult.
 \placefigure{fig5}

\subsubsection{Time Series Analysis}
To determine any  time variability in the selected  stars we chose 3 reference 
(comparison) stars (H89, H20, H139).
 To select and check stability of the reference stars, we chose a set of 
stars with  variances less than 0.01 mag over the observing interval. 
Power spectrums of the candidate reference stars were calculated and the ones 
showing most random power distribution were selected. H89 and H139 from this
set was used by Cohen et al. (2004) also. 
Hence we adopt these stars as the reference stars.
Two of these stars have F spectral type and H139  is a G0 star (Luhman
et al. 1998). The average magnitude of the selected
reference stars were used in the calculation of differential magnitudes.
Figure 5b  shows the mean magnitudes of the reference stars obtained
for each frame against the mean 
magnitude errors in measuring the magnitude of the reference stars 
for the data obtained during the observation period  of
five months. Each point is the relevant value for the selected stars 
under investigation. Scatter in the mean magnitude 
values of reference stars are due to different number
of frames used, changing between 800 and 1300, for each selected star.
The selection criteria results in a different number of frames for each
star. The mean of the reference stars scatter since the
star under question and the reference stars  
are extracted together from each frame.
Differential magnitudes of the selected
objects are calculated for each frame with the requirements: Selected
object and the reference stars are detected with an accuracy of 3 pixels;
magnitude error should be less than 0.2 mag; frame should have a PSF
FWHM $<6\farcs6$.
The mean magnitude of reference stars changes about 0.04 mag
while the deviation from the mean is 
about 0.001 in mean magnitude measurement errors.
 Each of the three reference stars displays a standard deviation of
the order 0.03 magnitude during 5 months of observation period. 


We used differential magnitudes in the time series analysis.
Differential photometry eliminates the atmospheric and other systematic
effects over hundred days of observations. These include seeing variations 
in a specific night and between observation days, and also pointing
variations of the order of $\sim0\fdeg3$ in large FOV ($1\fdeg8$).
 After the calculation of differential magnitudes 
We applied the Period98, Clean and Scargle methods 
to obtain the periodograms. The periodograms were calculated for 
the frequency range between 0 and 20 $d^{-1}$, 
so it was possible to search for periods as short
as few hours. The time series of each star  was searched
for periodicity by using the above mentioned three different period
search methods. 
 Most prominent period detected (whose confidence level is greater 
or equal to $5\sigma$) was given in Table 1 with 
its confidence level which is calculated in the way described in section 3.1. 
These periods are attributed to the rotation of star with large cool spots 
on its surface.  
The variance ($\sigma_{var}$) of the magnitude variations of each star during 
the observation interval is also shown in the Table. 

 \placetable{table1}

\subsubsection{Periodic Variables}
\begin{figure}[h]
 \includegraphics[clip=true,scale=0.4,angle=0]{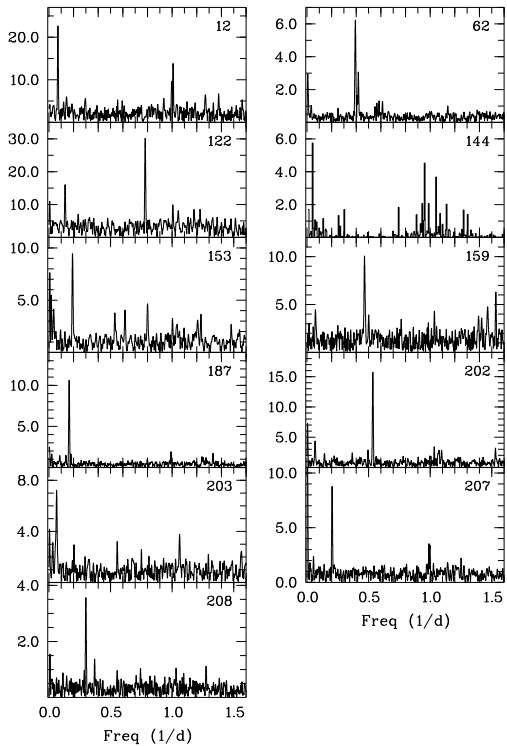}
 \figcaption{Amplitude spectra of newly detected periodic stars obtained
             by applying Clean method. Periods corresponding to the detected
             frequencies for each star  are given in Table 1. }\label{fig6}
 \end{figure}

\begin{figure}[h]
 \includegraphics[clip=true,scale=0.4,angle=0]{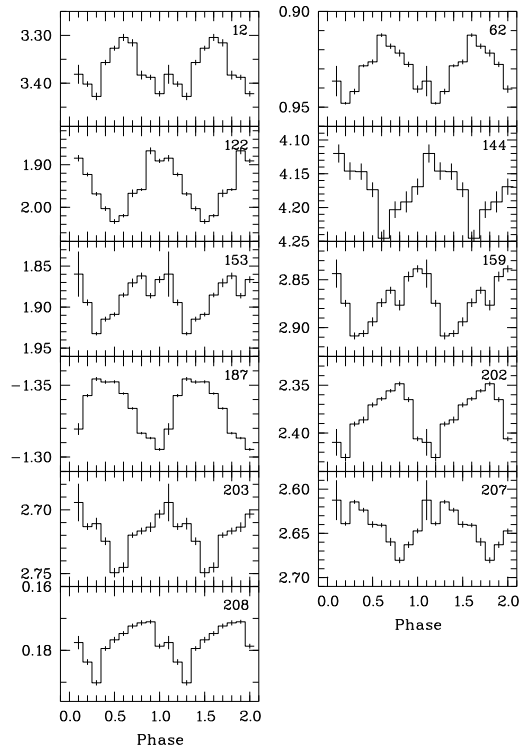}
 \figcaption{Phased light curves of newly detected periodic stars
             folded at the detected frequencies. Vertical axis is
             in magnitude units.}\label{fig7}
 \end{figure}

We found 35 stars  as periodic variables.   
Of the detected 35 periodic variables, 18 stars are new
periodic detections.
The rest of them whose HMW (Herbst et al. 2000) numbers are given in 
column 3 of Table 1 were studied also by Cohen et al. (2004).
The amplitude spectra of 11 newly detected periodic stars obtained
by applying Clean method
to the time series data of stars are shown in Figure 6. 
The rest of them which are stars 3, 20, 51, 71, 73, 143 and 173  are shown in 
Figure 8, 10 and 11.
 \placefigure{fig6}
 \placefigure{fig7}
Figure 7 shows phased light curves of the stars shown in Figure 6 at the 
detected frequencies. Binned phase diagrams are obtained by folding each time
series at the detected period. 
The amplitude of modulation for each star changes between 0.02 and 0.20 
magnitude.

\begin{figure}[h]
 \includegraphics[clip=true,scale=0.3,angle=270]{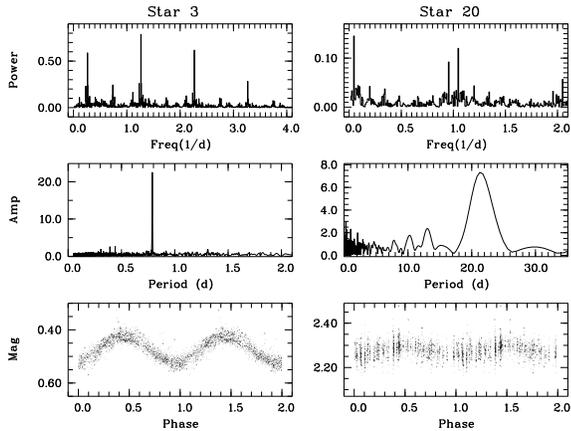}
 \figcaption{Star 3 and star 20: Stars having the shortest and longest periods in our sample.
            Upper panel shows the power spectra
            obtained by Scargle algoritm. Middle panel presents amplitude
            spectra obtained by Clean algoritm. 
            Lower panel is the phased light curve. }\label{fig8}
 \end{figure}

We display the power spectra  of  stars 3 and 20
in Figure 8, together with their phased light curves. 
 Stars 3 and 20 are the samples of stars having shortest and longest 
 periods  in our study. 
The other peaks seen in the top panel are the beat frequencies 
between the star's and Earth's rotation periods. In the middle panel 
cleaned dirty spectrum obtained by using Clean method is given.
 \placefigure{fig8}

\begin{figure}[h]
 \includegraphics[clip=true,scale=0.3,angle=270]{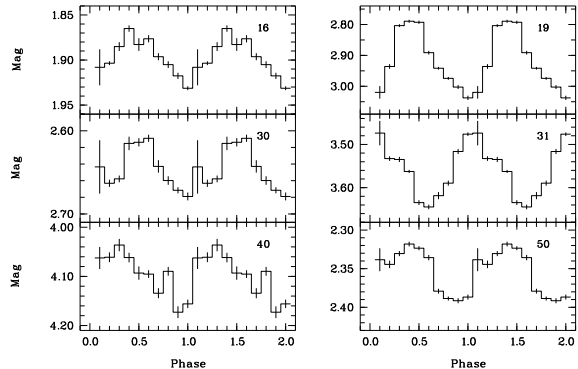}
 \figcaption{Differential light curves of 6 periodic stars in our
            sample which were also given by Cohen et al.
            Numbers for each star refers to HWM catalog.}\label{fig9}
 \end{figure}

Herbst et al. (2000) indicated that CTTSs were less likely
to exhibit periodic variations than WTTSs. 
Active accretion can prevent any rotational signature.
WTTSs are periodic stars. Their cool spots on the surface which are
stable for several months (Herbst et al. 2000, Cohen et al. 2004)
allow us to detect the rotation period. These cool spots are expected to 
be associated with magnetic fields. 
Detection of periodicity could be difficult if the spot pattern
and places of them change on a timescale of weeks. 
The periods determined by Cohen et al. and us 
are similar, that is they are similar with a maximum change
in period  by 1$\%$ except for star 114.   
We observed a period of 15.88 d for this star which
is greater about a half day compared to the value of Cohen et al.
They detected different periods for this star
in different seasons so they gave an average period for five seasons
which is 16.40 d. There is a period change of 3$\%$ for this star.
This can be related with the chosen Fourier step size which
gives a maximum error of 0.7 d.  Hence, the stability of rotation periods of
TT stars over long time scales is confirmed.
Cohen et al. remarks that the longest time that a spot configuration
can remain stable enough is between 0.5 and 1 yr.
 \placefigure{fig9}
In Figure 9, we plotted the light curves of 6 periodic stars in our
sample which were also plotted by Cohen et al. 
for the time interval between 1998 and 2003. 
These plots make stronger the remarks of Cohen et al. about the change
of light curve from one season to the other.

The spectral types of stars that we studied are between A0 and M4.
For spectral types earlier than late F type Chandra X-ray
detection fraction of the
cluster is less (Preibisch and Zinnecker 2001). Earlier spectral type
stars do not show intrinsic X-ray emission. 
 Therefore, star 94  did not show rotation 
period. For star 187 which is an A0 type star, we found a period of 6.097 d.
Since we do not expect a chromospheric activity that produce X-ray emission
from this spectral type,
no rotation period should be observed.
It can be explained in the way as Preibisch and Zinnecker (2001) explained;
that is this rotation period is due to a very close late type companion
it is not related with the star itself.

\subsubsection{CTT Variables}

\begin{figure}[h]
 \includegraphics[clip=true,scale=0.3,angle=270]{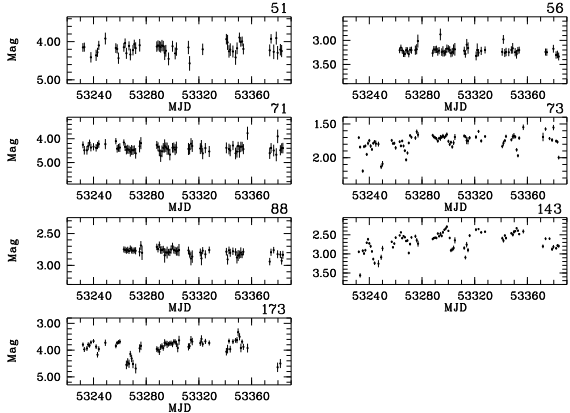}
 \figcaption{The daily averages of differential light curves of six CTTSs
             and star 173 which shows large variations in magnitude
             although classified as WTTS. }\label{fig10}
 \end{figure}

\begin{figure}[h]
 \includegraphics[clip=true,scale=0.42,angle=0]{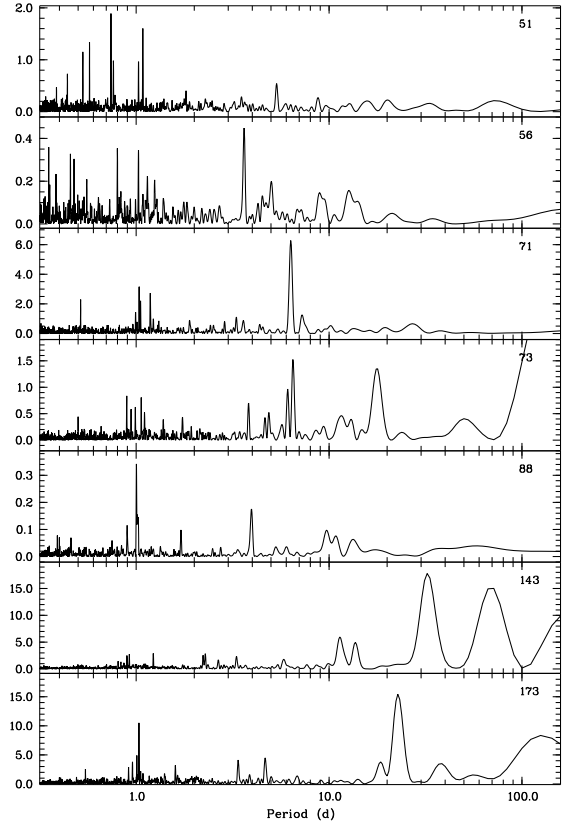}
 \figcaption{The amplitude spectra for the 
             CTTSs and star 173 of Figure 10. }\label{fig11}
 \end{figure}

 \placefigure{fig10}
 \placefigure{fig11}
The daily averages of differential light curves of CTTSs given in Table 1
are plotted in Figure 10. These are stars 51, 56, 71, 73, 88  and
143. The star 73 shows a magnitude variation of 0.7 magnitude.
The accretion activity is highly variable in time.
The continuous activity of this star with its deep minima is seen clearly from
the figure. If we think that minima shows the photospheric
luminosities then the increase in luminosity can be caused by
the accretion from a disk around the star. Herbig (1998) classifies this star
as CTTS (spectral type K0) while Luhmann et al. (1998) measurements of
H$_\alpha$ equivalent width indicates a WTTS. 
H$_\alpha$ emission seems to be time dependent in TT stars (Guenther $\&$
Emmerson 1997).
The star 143 also
shows similar variations like star 73, showing a magnitude variations of about
1.4 mag. Taking typical mass and radius of a K0 star we can calculate the mass accreted on to the star. If increase in luminosity is arising from the accretion from a disk which is comparable to that of gravitational contraction then a
change of 0.7 in magnitude corresponds to a mass accretion rate of
\.M$ \sim 10^{19}$ gr/s. This means that \.M is $\sim10^{-7}$
 \aaa/yr. 
 The other CTTSs are quiet that is variations in their magnitudes
 are small. 
 The amplitude  spectra of these stars are given in Figure 11.
Two of them (star 51 and 71) show 
rotational periodicities whose periods are given in
Table 1. Stars 143 and 73 may be in transition phase from
CTTS to WTTS as Herbst et al. (2000) and Cohen et al. (2004) suggested.
 They also suggested that the deep minima
 seen in the light curves of these stars could be
 caused by occultation events from dust clouds.
The maximum powers calculated are above 5$\sigma$ for these two stars
at the detected periods of 6.536 d (for star 73) and 32.28 d (for star 143)
which are probably rotation periods. 
Another star which shows activity in its differential light curve
 like stars 73 and 143
is star 173 (see Fig.10). This star shows magnitude variations of
 about 1.5 mag. Herbst et al. classifies this star as an active
non periodic WTTS since neither previous study gave the strength of 
hydrogen emission line (from which Herbst et al. inferred this line
 was weak). The amplitude spectrum of star 173 (Fig. 11) shows
a periodicity at 22.51 d with a 5$\sigma$ confidence level.
This star may also be thought as CTTS because of its high activity
similar to stars 73 and 143. 
Star 75 whose rotation period was calculated
 as 3.088 d was classified as U (unknown class) by
 Herbst et al. (2000). For this star,
 Luhman et al. (2003) gives the H$_\alpha$
equivalent width as 10\AA. It seems that this star has a
phase between CTT and WTT. Nevertheless, its light curve is rather quiet;
it does not show any activity  in its light
 curve as in the case of stars 73, 143 and 173.

\subsubsection{X-ray Variability and Rotational Periods}
In our search we mostly tried to find a period for the optical counterparts
of the Chandra sources which are classified as WTTSs.
The observed X-ray 
emission for WTTSs was explained as coronal emission by
 Preibisch and Zinnecker (2001) and related to the
stellar rotation.
All WTTSs as possible X-ray sources may not have a variability
during the time of observation they may be in their spot less
or changing spot pattern period
as in the case of stars 77, 103, 115, 148 and 163. To this list
we can include also star 17 and 188, since Luhman et al. (2003)
 gives the H$_\alpha$
equivalent width smaller than 10\AA~ for these stars.
 CTTSs have circumstellar accretion disk which could prevent
 the star to show a regular rotation. CTTSs were also detected as X-ray
 sources
(Preibisch and Zinnecker 2001). Detection frequency among the CTTSs is
45$\%$ while among WTTSs it is 73$\%$. They found no significant
difference between the X-ray properties of WTTSs and CTTSs.

Prebisch $\&$ Zinneker (2002) have shown the light curves (count rates) of
 the sources which shows strong variability during the 
Chandra observation. For most of these sources we found rotational
 periods whatever the character of variation of the count rates
(flare activity, rising or decaying of count rates). 

\begin{figure}[h]
 \includegraphics[clip=true,scale=0.3,angle=270]{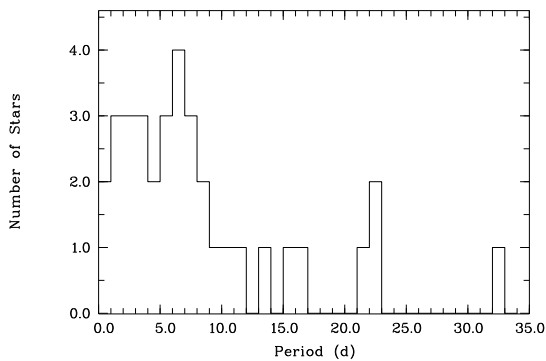}
 \figcaption{The distribution of rotation periods of sample stars 
          in IC 348 cluster for spectral types earlier than M4}\label{fig12}
 \end{figure}

 \placefigure{fig12}
In Figure 12, we plot the distribution of rotation
 periods in IC 348 cluster using our
 sample sources for spectral types earlier than M4. 
The number of stars with slow rotation is less.
Cohen et al. (2004) mentioned about the absence of periods 
shorter than 1 day and deficiency of periods between 4 and 5
days. They said these characteristics were also shared
 with the period distributions
of the Orion Nebula Cluster and Taurus.
 We have two stars whose periodicity is shorter than 1 day.
The plot of IC 348 cluster star distribution is similar to Cohen et al.'s
except we have periods shorter than 1 day and longer than 16 days.

Plot of rotational period versus spectral type is shown in Figure 13.
There is an increase in period towards the later spectral types.
Stars whose spectral types later than K3 have wide range of periods
 between 0.74 and 32 d. However, an overall gradual increase  can not
be ruled out.
Whereas G and early K dwarfs have smaller rotation  periods with a mean 
value of $\sim3.7$ days.
In Figure 14, we investigate how TTS's rotation
is related to the chromospheric and coronal activity.
X-ray luminosities of the sample stars given by \citet{pre02} were plotted
against the rotational period.
Bouvier (1990) proposed that the correlation between X-ray fluxes and
 rotational periods of TTSs was caused by a solar type dynamo which is 
responsible for the chromosperic and coronal activity of stars as it is 
in active dwarfs.
Despite the large scatter in the data, there is a trend toward decreasing
X-ray luminosity as the rotation period increases.
On the other hand stars with periods $<$4 d have 
an average X-ray luminosity of $\sim 2\times10^{30}$ erg/sec 
with a large scatter.
As the stars rotate faster their chromospheric and coronal activity increases.
Rotation seems to be an important parameter which influences the level of
X-ray emission of stars.
We note that WTTSs and CTTSs exhibit similar X-ray luminosities 
at any rotational period.
We conclude that X-ray luminosities of TTSs in IC 348 cluster depend on
 rotation.

\begin{figure}[h]
 \includegraphics[clip=true,scale=0.3,angle=270]{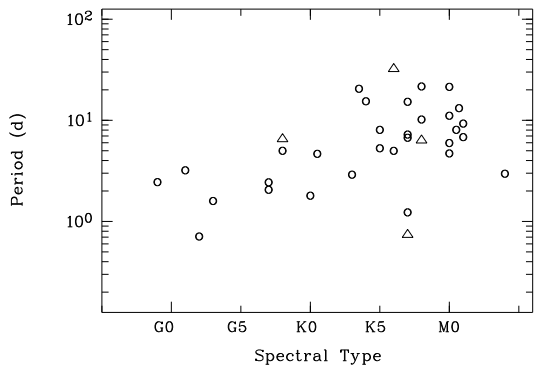}
 \figcaption{The distribution of rotation periods of sample stars
          in IC 348 cluster. 
           Open circles denote WTTSs and triangles are CTTSs.
           }\label{fig13}
 \end{figure}

 \placefigure{fig13}

\begin{figure}[h]
 \includegraphics[clip=true,scale=0.3,angle=270]{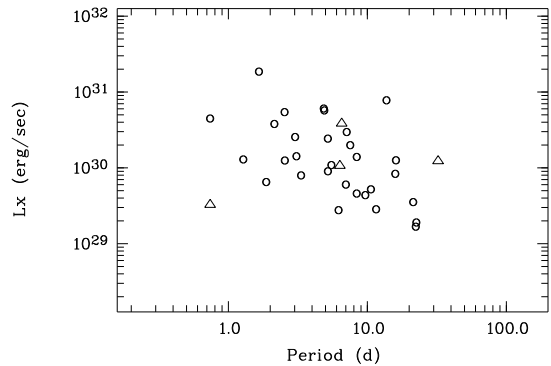}
 \figcaption{The distribution of X-ray luminosity  of sample stars
          in IC 348 cluster as a function of rotational periods.
          Open circles denote WTTSs and triangles are CTTSs.}\label{fig14}
 \end{figure}

 \placefigure{fig14}

\section{Summary}

The main results of our analysis of the ROTSE observations of IC 348 
cluster can be summarized as follows:

We have 5 months of continuous data of this cluster.
In the time series analysis of the stars for the frequency range
 between 0 and 20 $d^{-1}$ we did not find any periodicity shorter than 
0.7 d. Only for the star H254 we confirmed the  
\ee pulsation period of 3.24 hr. The other frequencies detected
0.157, 0.283 and 0.931 $d^{-1}$
by Ripepi et al. (2002) for  H254 were not present in our light curve.

We found 35 stars as rotationally periodic stars whose rotation
periods change between 0.74 and 32.3 d. 18 of them were  newly 
detected periodic stars. 8 of the 18 stars (stars  20, 62, 73, 122,
143, 153, 159, 173) were also studied by 
Cohen et al. but they did not give any period for
these stars. That can be due to the unstable spot patterns
during their observation periods. Perhaps the observation duration was
not enough to determine the periods.
Cohen et al. noted that stars may remain spotted but the spot pattern
evolves such that a period can not be determined over 6 consecutive months
of observation. Since we detected the periods of these stars it is
probably related with the number of data points used in the time series. 
If the size of the spot is small it can be difficult to detect the period. 

Most of the stars whose periods were detected were WTTSs.
The periods of non accreting WTTSs are easily detected.
There were 7 non periodic WTTSs in our analysis. This may be due
to a changing spot pattern or spot less period of the star
during the observation.
For one of them (star 103) Cohen et al. gives a period of
2.237 d which is an average over all of five seasons
they studied. They could not find periodicity for two seasons 
for this star.
The number of CTTSs that we study is less than WTTSs.
For the 4 CTTSs (51, 71, 73, 143) we detected rotation period.
It would not be possible to detect the periods if the disk
prevents the detection of rotational variability.
Prebisch $\&$ Zinneker (2002) noted that M type stars without any
circumstellar material can show H$_\alpha$ emission.
They may not have an accretion disk.
H$_\alpha$ emission can also be time dependent in TT stars.
These 4 stars may be seen as CTTS at the time of measurement of
H$_\alpha$ emission, but at another time H$_\alpha$ emission
 may be less and they may appear as WTTS. 

The rotational periods found in this study are similar with those of 
Cohen et al. with a maximum change of 1$\%$ in period. 
Small changes in the rotational periods indicate a rigid rotation.
Rotation periods seems to be stable on a timescale of 
$\sim$6 yr in this cluster 
when evaluated together with the results of Cohen et al.

We found an inverse correlation between X-ray luminosity and the rotational
period in our sample of late type TTSs.
X-ray luminosity decreases as the  stars
rotate slower. WTTSs and CTTSs bahave similar in X-ray activity
at any rotational period. The dispersion in rotational periods at a given
spectral type results in a  dispersion in X-ray luminosity.

\acknowledgments

We thank the referee Prof. Kevin Luhman, for a careful reading and 
valuable comments.
This project utilizes data obtained by the Robotic Optical Transient Search
Experiment.  ROTSE is a collaboration of Lawrence Livermore National Lab, 
Los Alamos National Lab, and the University of Michigan
(www.umich.edu/$\sim$rotse).
We thank the Turkish National Observatory of TUB\.ITAK
for running the optical facilities. This research has made use of the 
SIMBAD database, operated at CDS, Strasbourg, France. 
Special thanks to Tuncay \"Oz{\i}\c{s}{\i}k from TUG 
who keeps hands on ROTSEIIId.

\clearpage
\begin{deluxetable}{ccccccrc}
\tablecolumns{8}
\tablewidth{0pc}
\tablecaption{Properties of Sample IC 348 Field Stars \label{table1}}
\tablehead{
\colhead{Star}          & 
\colhead{OTHER IDs}      & 
\colhead{HMW}            & 
\colhead{SP Type \tablenotemark{a}}       & 
\colhead{$\sigma_{var}$} & 
\colhead{Type \tablenotemark{b}}          &
\colhead{Period}         &
\colhead{Conf \tablenotemark{c}}} 
\startdata
 3   & LRL22  &\nodata &  G2    & 0.063  &\nodata & 0.789  & \\
 8   & LRL47  &\nodata &  K0.5  & 0.039  &\nodata & 4.857  & \\
 12  & LRL87  &\nodata &  M0.7  & 0.040  & (W)    & 13.73  & \\
 17  & H39    & 62     &  M1.5  & 0.178  & U(W)   &\nodata & $<3\sigma$\\
 20  & H43    & 24     &  K3.5  & 0.004  & U      & 21.37  & \\
 26  & H63    & 52     &  K8    & 0.050  & W      & 10.61  & \\
 32  & H70    & 26     &  K3    & 0.008  & W      & 3.021  & \\
 49  & H93    & 60     &  M1    & 0.148  & W      & 7.102  & \\
 51  & H94    & 59     &  K7    & 0.110  & C      & 0.739  & $ 5\sigma$\\
 52  & H95    & 40     &  K5    & 0.052  & W      & 8.382  & $ 5\sigma$\\
 56  & LRL100 & 34     &  M2    & 0.107  & C      &\nodata & $<3\sigma$\\
 61  & H102   & 39     &  M1    & 0.082  & W      & 9.667  & \\
 62  & H103   &  9     &  F9Ve  & 0.023  & G      & 2.548  & \\
 70  & LRL64  & 19     &  M0.5  & 0.052  & U(W)   & 8.385  & \\
 71  & LRL60  &133     &  K8    & 0.108  & C      & 6.340  & \\
 73  & H114   & 20     &  G8    & 0.249  & C      & 6.536\tablenotemark{d}& \\
 75  & LRL62  & 31     &  M4    & 0.149  & U      & 3.088  & \\
 77  & H116   & 21     &  M0    & 0.035  & W      &\nodata & $<3\sigma$\\
 83  & H121   & 41     &  K7    & 0.009  & W      & 7.004  & \\
 88  & H124   & 75     &  K6    & 0.044  & C      &\nodata & $<3\sigma$\\
 94  & H252   & 10     &  A2    & 0.004  & E      &\nodata & $<3\sigma$\\
 103 & H137   & 12     &  K2    & 0.048  & W      &\nodata & $<3\sigma$\\
 114 & H148   & 44     &  K7    & 0.009  & W      & 15.88  & $ 5\sigma$\\
 115 & H150   & 54     &  M1    & 0.015  & W      &\nodata & $<3\sigma$\\
 119 & H155   & 50     &  K5    & 0.044  & W      & 5.509  & \\
 122 & H157   &143     &  K7    & 0.024  & W      & 1.280  & \\
 133 & H166   & 11     &  G3    & 0.008  & G      & 1.659  & $ 5\sigma$\\
 137 & H171   & 49     &  M0    & 0.042  & W      & 6.207  & \\
 143 & LRL37  & 23     &  K6    & 0.169  & C      &32.28\tablenotemark{d}& \\
 144 & LRL144 &\nodata &  M0    & 0.020  & (W)    & 22.31  & $ 5\sigma$\\
 145 & H178   & 16     &  K6    & 0.039  & W      & 5.197  & \\
 146 & H179   & 30     &  K7    & 0.108  & W      & 7.532  & \\
 148 & H181   & 46     &  M2    & 0.240  & W      &\nodata & $<3\sigma$\\
 151 & H184   & 14     &  G7    & 0.038  & G      & 2.537  & \\
 153 & H187   & 15     &  G8    & 0.063  & G      & 5.203  & \\
 159 & LRl59  & 18     &  G7    & 0.026  & G      & 2.142  & \\
 163 & H198   & 82     &  M3    & 0.142  & W      &\nodata & $<3\sigma$\\
 166 & H205   & 51     &  M0    & 0.005  & W      & 11.56  & \\
 173 & H214   & 56     &  K8    & 0.647  & W      &22.51\tablenotemark{d}& \\
 176 & LRL15  & 22     &  M0.5  & 0.039  & C      &\nodata & $<3\sigma$\\
 187 & LRL3   &\nodata &  A0    & 0.004  &\nodata & 6.097  & \\
 188 & LRL101 &\nodata &  M3.2  & 0.008  & (W)    &\nodata & $<3\sigma$\\
 202 & LRL79  &\nodata &  K0    & 0.004  &\nodata & 1.872  & \\
 203 & LRL39  &\nodata &  K4    & 0.048  &\nodata & 16.06  & \\
 207 & LRL1937&\nodata &  M0    & 0.043  & (W)    & 4.900  & \\
 208 & LRL20  &\nodata &  G1    & 0.005  &\nodata & 3.335  & \\
\enddata
\tablecomments{Star  numbers are from the catalog (CXOPZ) given by \citet{pre01}.
          OTHER IDs starting with H and L are from 
          \citet{her98} and \citet{luh98} respectively. 
          HMW numbers are from \citet{her00}.
          Units of $\sigma_{var}$ is magnitude and period is in days. }
\tablenotetext{a}{Spectral types are from \citet{luh98}
                 and Centre de Donn\'ees astronomiques de Strasbourg 
                 SIMBAD database.}
\tablenotetext{b}{Spectral category from \citet{her00}; 
                  W: Weak line TTs; C: Classical TTs; U: Unknown;
                  G: G type; E: Early type;
                  (W): WTTS assigned in this study considering the 
              H$_\alpha$ equivalent width given by Luhman et al. (2003).}
\tablenotetext{c}{Confidence level of the calculated period: Blank if
                  more than $5\sigma$. $<3\sigma$ means either
                  no periodicity is detected or the confidence level
                  of the detected period is small.}
\tablenotetext{d}{See text.}
\end{deluxetable}

 \end{document}